\documentclass[12pt]{article}

\usepackage{geometry}
\usepackage{booktabs}
\usepackage[parfill]{parskip} 
\usepackage{graphicx}
\usepackage{amsmath, amssymb, amsthm, bbm}
\usepackage[font=small,labelfont=bf]{caption, subcaption}
\usepackage[T1]{fontenc}
\usepackage{setspace}
\usepackage[loose,nice]{units}
\usepackage{array}
\usepackage[super]{nth}
\usepackage{graphicx}
\usepackage{float}
\usepackage{varioref}
\usepackage[unicode=true,colorlinks=true,urlcolor=blue,citecolor=black,linkcolor=black]{hyperref}
\usepackage{cleveref}
\usepackage{subcaption}
\usepackage{mathtools}
\usepackage[all]{nowidow}
\usepackage{wrapfig}
\usepackage{pdfpages}
\usepackage{authblk}
\usepackage[utf8x]{inputenc}
\usepackage[english]{babel}
\usepackage[title,page]{appendix}
\usepackage{chngcntr}
\usepackage[shortcuts]{extdash}

\usepackage{titlesec}
\titlespacing*{\section}
{0pt}{2ex plus 1ex minus .2ex}{2ex plus .2ex}
\titlespacing*{\subsection}
{0pt}{1ex plus 1ex minus .2ex}{1ex plus .2ex}
\titlespacing*{\paragraph}
{0pt}{1ex plus 1ex minus .2ex}{1ex plus .2ex}

\usepackage{newtxtext} 
\usepackage{amsmath}
\usepackage[bigdelims]{newtxmath}
\usepackage[T1]{fontenc}
\usepackage{textcomp}
\usepackage{hyperref}
\usepackage{xcolor}

\usepackage[round,colon,authoryear]{natbib}

\DeclareMathOperator*{\E}{{\rm I\kern-.3em E}}

\let\vec\mathbf

\title{
Bayesian estimation of the number of significant principal components for cultural data}

\author[$^{1,2}$]{Joshua C. Macdonald}
\author[$^3$]{Javier Blanco-Portillo}
\author[$^{3,*}$]{Marcus W. Feldman}
\author[$^{1,4,*}$]{Yoav Ram}

\affil[1]{School of Zoology, Faculty of Life Sciences, Tel Aviv University, Tel Aviv, Israel}
\affil[2]{Current Address: Department of International Health, Bloomberg School of Public Health, Johns Hopkins University, Baltimore, MD, USA}
\affil[3]{Department of Biology, Stanford University, Stanford, CA, USA}
\affil[4]{Safra Center for Bioinformatics, Tel Aviv University, Tel Aviv, Israel}
\affil[*]{Corresponding author, MWF:~mfeldman@stanford.edu, ~YR:~yoavram@tauex.tau.ac.il}

\begin{document}
\maketitle
\newpage 
\begin{abstract}
Principal component analysis (PCA) is often used to analyze multivariate data together with cluster analysis, which depends on the number of principal components used. It is therefore
important to determine the number of significant principal components (PCs) extracted from a data set. Here we use a variational Bayesian version of classical PCA, to develop a new
method for estimating the number of significant PCs in contexts where the number of samples is of a similar to or greater than the number of features. This eliminates guesswork and potential bias in manually determining the number of principal components and
avoids overestimation of variance by filtering noise. This framework can be applied to datasets of different shapes (number of rows and columns), different data types (binary, ordinal, categorical,
continuous), and with noisy and missing data. Therefore, it is especially useful for data with arbitrary encodings and similar numbers of rows and columns, such as cultural, ecological, morphological,
and behavioral datasets. We tested our method on both synthetic data and empirical datasets and found that it may underestimate but not overestimate the number of principal components for
the synthetic data. A small number of components was found for each empirical dataset. These results suggest that it is broadly applicable across the life sciences.\end{abstract}\newpage 

\section*{Introduction}

Principal component analysis (PCA) is commonly applied to multivariate datasets to reduce the number of covariates analyzed to a tractable number and is a ubiquitous approach to dimension reduction.
However, PCA requires the practitioner to manually choose the number of dimensions in the reduced principal components space, which can be interpreted as the number of significant groups of covariates in a dataset, or the number of underlying `factors'.
In some cases a single principal component explains most of the variance in a dataset \citep[e.g.,][]{turchin2018quantitative}. But in many cases the number of principal components is not immediately clear and can potentially bias downstream analysis \citep{patterson2006population,beheim2021treatment}.

Methods to determine this number have developed in different scientific fields, 
such as permutation bootstrapping of the principal components \citep{vasco2012permutation,karin2018tradeoffs,camargo2022pcatest}, exploratory factor analysis, which clusters co-varying features (columns) over permutations of the data \citep{courtney2013determining}, or selecting all the eigenvalues of the covariance matrix that exceed one \citep[\emph{Kaiser's criterion}, ][]{kaiser1958varimax}. In some fields, theory-based methods have been developed for determining the number of significant principal components \citep[e.g., in population genetics,][]{patterson2006population}.
However, these methods usually make assumptions about the data that restrict their broader application.

Here, we develop a new method for estimating the number of significant principal components that involves a Bayesian version of the classical PCA algorithm, \emph{variational Bayesian principle component analysis} \citep[VBPCA,][]{ilin2010practical} and uses the posterior distributions inferred by VBPCA.
In the following, we describe VBPCA and our method, and then demonstrate its application to synthetic and empirical data.

\section*{Methods}
In the following, uppercase symbols are for matrices ($X$), bold symbols are for vectors ($\vec x$), and plain symbols are for scalars ($x$).
The normal (Gaussian) distribution with expectation $\mu$ and variance $\sigma^2$ is written as $\mathcal{N}(\mu, \sigma^2)$.

\subsection*{Variational Bayesian principal component analysis}
In the standard PCA algorithm, the linear projection of an $n \times p$ data matrix $X$ onto a principal component space of dimension $q \leq \min\{n,p\}$ can be written as
\begin{equation}
    X^T \approx AY + M \;,
\label{eq:PCA}    
\end{equation}
where $A$ is a $p\times q$ loadings matrix, $Y$ is a $q\times n$ matrix whose rows are the $q$ principal components, and $M$ is bias matrix composed of $n$ copies of $\vec m$, a $p\times 1$ bias vector.

There are several probabilistic formulations of PCA \citep[e.g.,][]{tipping1999probabilistic, PPCAcode}.
Here, we focus on \emph{variational Bayesian PCA} \citep[VBPCA,][]{ilin2010practical}, which can be used to simultaneously impute missing values and filter noise in $X$.
Moreover, VBPCA estimates posterior distributions for the data entries $x_{i,j}$, which are the focus of our interest.
We use an implementation of VBPCA written in \emph{Matlab} by~\citet{VBPCAcode}.
Here we provide the details essential for our purposes, see \citet{ilin2010practical} for a the full details.

The cost function of the VBPCA algorithm, given $q$ principal components, is the squared error for reconstruction, $\vert \vert X - AY - M \vert \vert_F^2$; that is, the Frobenius norm of the difference between the original data matrix $X^T$ and the reconstructed data matrix $AY+M$. 
The VBPCA algorithm uses a combination of expectation-maximization (EM) and variational inference. This is an iterative algorithm, which we terminate after 80 iterations, as we observed that the cost function stopped decreasing in all cases examined. In practice 30 iterations are usually sufficient \citep[see Fig. S21 of ][]{Macdonald2023results}. 

VBPCA estimates $\hat{A}$ and $\Sigma_A$, $\hat{Y}$ and $\Sigma_Y$, and $\hat{\vec m}$ and $\Sigma_{\vec m}$, where $\hat{\cdot}$ and $\Sigma_{\cdot}$ are the mean and the covariance matrix of each parameter.
We use these parameter estimates to estimate the mean $\hat{X}$ and covariance $\Sigma_X$ of the data reconstruction matrix (\cref{eq:PCA}) using

\begin{equation}
    \hat{X}^T = \hat{A}\hat{Y} + \hat{M} \;,  
    \label{recon}
\end{equation}
and
\begin{equation}
    (\Sigma_X)_{i,j} = \hat{A}_j \Sigma_{Y_i} \hat{A}_j^T + \hat{Y}_i^T\Sigma_{A_j}\hat{Y}_i + \sum_{\ell=1}^q\sum_{k=1}^q [\Sigma_{Y_i}\odot \Sigma_{A_j}]_{\ell,k} + \Sigma_{\vec{m},j} \;,
\label{postvar}
\end{equation}
where $A$, $\hat{A}$ are indexed by column $j$ and $Y$, $\hat{Y}$ are indexed by row $i$.

\subsection*{New method for estimating the number of significant principal components}
To estimate the number of significant components in a dataset, we extend a population-genetic method developed by
\cite{patterson2006population} that determines the effective number of features (columns) in a dataset.
According to \cite{patterson2006population} given a data matrix, $X$ with $n$ rows and $p$ columns and $C = XX^T$, 
If we assume that the entries of the data matrix $X$ are independently and normally distributed and that $X$ has many more columns than rows, $p >> n$, then the (normalized) eigenvalues of $C$ will approximately follow a Tracy-Widom (TW) distribution. This allows the number of significant non-random eigenvalues to be determined using hypothesis testing \citep{patterson2006population}.

However, while theses assumptions are met in genetics datasets these assumptions do not hold for  cultural/anthropological datasets, which usually have similar numbers of rows and columns. 
Therefore, the eigenvalues of $C$ are not expected to follow a TW distribution.

Here we propose a method to circumvent this problem. We leverage the posterior distribution of the VBPCA reconstructed data matrix (\cref{recon,postvar}) to develop a test similar to the TW test of \cite{patterson2006population} that makes no assumption on the number of rows and columns of the data matrix or the distributions of the eigenvalues themselves.
Thus, instead of assuming the normalized eigenvalues of $C$ are samples from a TW distribution, we produce $N$ posterior samples $\{\tilde{\ell}_{k,r}\}_{k=1}^{N}$ from the theoretical distribution $\tilde{L}_r$ of each normalized eigenvalue using the posterior distributions estimated by VBPCA. These posterior samples can then be compared to the eigenvalue to test if that eigenvalue is larger than expected by chance. We now provide the full details of this algorithm.

First, we center the data matrix $X$ by subtracting from each column its mean.
Next, we find the number of principal components $q$ that best reconstructs the data using VBPCA, namely, that minimizes the mean squared error given $\tilde q$ principal components
\begin{equation}
q = \textrm{ arg min}_{\tilde q}  \vert\vert X- \hat{X}_{\tilde{q}} \vert \vert_F^2,
\label{cost_func} 
\end{equation}
where $X$ is the data matrix and $\hat{X}_{\tilde{q}}$ is the mean of the VBPCA posterior reconstructed data matrix (\cref{recon}) using $\tilde{q}$ principal components
(note that $\tilde{q}$ is the only hyper-parameter of the VBPCA algorithm and that $2 \leq \tilde{q} \leq \min\{n,p\}$).
After finding $q$, we can use the posterior distribution of the associated data matrix  with mean $\hat{X}$ and covariance $\Sigma_{X}$ (\cref{recon,postvar}), omitting the subscript $q$ for simplicity.

The eigenvalues of $\hat{C} = \hat{X}\hat{X}^T$   
are $\lambda_1 > \lambda_2 > \ldots > \lambda_{q} > 0$.
We aim to find the number of significant eigenvalues $w$ where $\lambda_r$ is a significant eigenvalue if and only if $r \le w$.
To find $w$, we normalize each eigenvalue, ${\ell}_r = \frac{(q-r)\lambda_r}{\sum_{s=r}^{q-1}\lambda_s}$ for $r=1,\ldots,q$ (note that ${\ell}_q=0$).
Next, we construct a theoretical distribution for the normalized eigenvalues under the null hypothesis that data entries are independent, that is, $x_{i,j} \sim \mathcal{N}(\hat{x}_{i,j}, (\Sigma_X)_{i,j})$, where the arguments of this normal distribution are the $i,j$ entries of the mean $\hat{X}$ and the covariance matrix $\Sigma_X$ of the posterior distribution of the VBPCA reconstructed matrix (\cref{recon,postvar}).
To do this, we sample $N$ data matrices from this null distribution, $\{\tilde{X}_k\}_{k=1}^{N}$.
We chose $N=2,000$, but this value can be changed.
For each data matrix sample $\tilde{X}_k$ we compute the eigenvalues of $\tilde{C}_k = \tilde{X}_k\tilde{X}_k^T$  order them in descending magnitude, $\tilde{\lambda}_{k,1} > \tilde{\lambda}_{k,2} > \dots > \tilde{\lambda}_{k,q} > 0$, and normalize them, $\tilde{\ell}_{k,r} = \frac{(q-r)\tilde{\lambda}_{k,r}}{\sum_{s=r}^{q-1}\tilde{\lambda}_{k,s}}$.
We now have $N$ samples $\{\tilde{\ell}_{k,r}\}_{k=1}^N$ from the theoretical distribution of the $r$-th normalized eigenvalue, $\tilde{L}_r$.
Finally, we calculate a p-value for each eigenvalue, $p_r$, $r = 1, \dots, q$, as the fraction of samples $\tilde{\ell}_{k,r}$ (out of $N$) larger than the normalized eigenvalue ${\ell}_r$, that is, 
$p_r = \frac{1}{N} \sum_{k=1}^{N}{\mathbbm{1}_{\tilde{\ell}_{k,r} > \ell_{r}}}$. 
We then adjust the p-value using the sequential Holm-Bonferroni method to correct for the familywise error rate \citep{abdi2010holm}.
After computing $p_{r}$ we continue to compute $p_{r+1}$ only if $p_r<\alpha$, where $\alpha$ is the significance level, e.g., $\alpha=0.05$.
We then set $w$ as the number of significant eigenvalues for which $p_1,\dots p_w < \alpha$.

\subsection*{Generation of synthetic data} 
To validate our algorithm for estimating the number of significant eigenvalues, we generated 200 synthetic datasets according to the distributions of the principal component matrix $Y$ and transition matrix $A$ above (\cref{recon}), with the following assumptions on their covariance matrices, to generate data with an expected number of significant principal components. 
Suppose that we have $n$ samples and $p$ features and we expect $w$ significant PCs, then the covariance matrices for $Y$ and $A$  are set to be 
\begin{equation}
    \begin{split}
        (\Sigma_Y)_{ij} &= 
        \begin{cases}
        0.001,  & i = j < p - w \;, \\
        1-0.03\left(j - (p - w) \right), & i=j \ge p - w \;, \\ 
                0, & i \ne j \;,
        \end{cases} \\
        (\Sigma_A)_{ij} &= 
        \begin{cases}
        0.00015, & i=j \;, \\
        0, & i \ne j \;.
        \end{cases}
    \end{split}
    \label{eq:vars}
\end{equation}

We examined two scenarios: Scenario (i) with $n$=150 rows (samples), and either $q$=15, 30, 35, 40, 45, 50, or 55 columns (features), and with either $w$=2, 4, 6, or 8 significant principal components; and Scenario (ii) with $p$=150 columns, either $n$=15, 30, 35, 40, 45, 50, or 55 rows; and with either $w$=2, 4, 6, or 8 significant principal components. 

\subsection*{Empirical data} 

We apply our method to six cross-cultural datasets described below. Teams of subject-matter experts encoded the cultural variables in the datasets based on previous ethnographies.  \Cref{tab:results} summarizes the statistical properties of these datasets resulting from our analysis.

\paragraph{Ethnographic Atlas.} The Ethnographic Atlas (EA) \citep{murdock1967ethnographic,dplace} includes 1,291 cultural groups from around the world \citep{murdock1967ethnographic,dplace} and many cultural variables that are divided into the categories of \emph{Subsistence}, \emph{Labor}, \emph{Community organization}, and \emph{Kinship practices}, among others. We focus on 69 EA features (Supplementary Data) and the $n$=130 cultural groups of people that speak an Austronesian language. We split these data into two datasets based upon the broad classifications \citep[see][]{Macdonald2023results}: \emph{Social Organization} (dataset Soc) and  \emph{Subsistence} (dataset Sub). 

\paragraph{Pulotu.} Pulotu \citep{pulotu,dplace} includes $n$=137 Austronesian cultural groups and was compiled for studying religion in the Austronesian people \citep{pulotu,dplace}. It also contains data on other aspects of culture and several geographical and historical features, such as island type and the presence or absence of pre-Austronesian populations. We focus on 36 features from Pulotu that are informative on traditional/indigenous cultural traits and conflict within and between groups and are missing no more than 50\% of their values. We split these data into two datasets \citep[see][]{Macdonald2023results}: \emph{Religion} (dataset Rel) and \emph{Cultural Interaction} (dataset Inter). 

\paragraph{Binford Hunter-Gatherer.}
Binford Hunter-Gatherer dataset (Dataset Bin) \citep{dplace,binford2019constructing} contains 36 categorial and ordinal features describing subsistence and kinship practices of $n$=339 globally distributed cultural groups as well as four continuous traits that quantify differences in body size. Approximately two thirds of the groups are also in EA, but the dataset includes cultural groups from Australia and North America that are absent in EA.

\paragraph{AVONET.}
The AVONET dataset \citep{tobias2022avonet} is a comprehensive functional trait dataset of all birds that includes six ecological variables, 11 continuous morphological traits, and information on range size and location. We focus on the morphological and ecological features, which are a mixture of continuous and discrete features (Dataset Bird) .

\paragraph{Seshat.} The Seshat dataset  (Dataset Ses) has $p$=9 features describing cultural complexity across the globe at more than 400 different time points. Seshat is used as a baseline empirical dataset since the  number of significant principal components is small and clear, $w$=1 \citep{turchin2018quantitative}. 

\paragraph{Encoding and pre-processing.}
The features in these datasets may be  continuous, categorical, or ordinal. An example from Pulotu is \emph{degree of external conflict}, which takes integer values between 1 (\emph{frequent external conflict}) and 4 (\emph{absence of external conflict}).  
Subject matter experts assigned multiple categories to each feature in the original datasets (eg. \emph{Sex differences: fishing} from EA which is categorized under labor, gender, and economy).  
These features are often on an arbitrary scale: discrete features have between two to 11 possible values. Some features are also missing a majority of their entries, and we removed any columns missing more than 50\% of their values.
We then binarized the discrete features using `one-hot encoding', a common practice in multivariate statistics and machine learning for representing categorical data (i.e., `dummy variables').
Therefore, the number of columns $p$ in \Cref{tab:results} is larger than the number of features described in the previous paragraphs.
We then rescaled the features by subtracting the mean from each column. For continuous features, we also divided column entries by their standard deviation.


\section*{Results}

\subsection*{Estimating the number of significant principal components: synthetic data.}
Applying our method to synthetic datasets, we find that the method may underestimate the number of significant PCs when the number of rows, $n$, or columns, $p$, is low and the actual number of significant PCs is high (\Cref{fig:testval}).
Notably, even for eight significant PCs, a moderate increase in the number of rows (samples) or columns (features) appears sufficient to accurately estimate the number of significant PCs.
For example, for a $150 \times 30$ (or $30 \times 150$) data matrix with eight significant PCs, our method estimates 5--6 significant PCs, but with 55 columns (or rows), our method correctly estimates the number of significant PCs (\Cref{fig:testval}).  

\subsection*{Estimating the number of significant principal components: empirical data.}

First, the number of PCs needed for optimal data reconstruction (i.e., minimization of the VBPCA cost function, \cref{cost_func}) was between 13 and 49 (Table \ref{tab:results}, \Cref{fig:archnum}A). 
The proportion of variance explained for each principal component is sometimes used by practitioners to determine the number of significant principal components by choosing the number of PCs for which the proportion of variance significantly drops. 
This procedure is sometimes called `Cattell's scree test' \citep[][]{cattell1966scree}. 
Seshat \citep{turchin2018quantitative} is a clear example (\Cref{fig:archnum}B).
However, this procedure is not recommended by \cite{courtney2013determining}, because ``it suffers from ambiguity and subjectivity when there is no clear break or hinge in the depicted eigenvalues.''
Indeed, in many cases, this criterion is hard to apply: except for Seshat, it is unclear how to determine the number of principal components by looking at the scree plot (\Cref{fig:archnum}B).

\begin{table}[h!]
\caption{Summary of results for the considered datasets.}
\centering 
\begin{tabular}{lccccccc}
\toprule
& Soc & Sub & Rel & Inter & Bin & Bird  & Ses \\
\midrule
Rows, $n$ & 130 & 130 & 137 & 137 & 339 & 9993 &  414\\
Columns, $p$ & 235 & 115 & 79 & 18 & 67 & 42 & 9\\
Reconstruction PCs, $q$ & 48 & 36 & 30 &  12 & 41 & 31  & 8\\
Significant PCs, $w$ & 4 & 3 & 3 & 1 & 3 & 2 & 1 \\
P-value, $w$ & 0.007 & 0.006 & 0.013 & 0.015 & 0.016 & 0.005  &  < 0.0005  \\
P-value, $w+1$ & 0.09 & 0.053 & 0.13 &  0.47 & 0.45 & 0.07 & 1.0 \\
Cumulative variance, $w$ & 0.365 & 0.47 & 0.292 & 0.251 & 0.43 & 0.411 & 0.888  \\
Cumulative variance, $w+1$ & 0.412 & 0.544 & 0.358 & 0.407 & 0.506 & 0.535 & 0.933  \\
\bottomrule
\end{tabular}
\label{tab:results}
\end{table}


The number of significant principal components, $w$, estimated by our new method is between 1 and 4 (\Cref{fig:archnum}C, \Cref{tab:results}), much lower than the number required for optimal data reconstruction, $q$. 
Notably, the p-values consistently "jump," 
the ratio of the first p-value above $\alpha$ and the last p-value below $\alpha$ is large in all considered datasets (for Seshat we can only get a lower bound for because none of the 2,000 sampled eigenvalues was greater than the empirical value.)
These "jumps" may help to remove ambiguity in the choice of the significance level, $\alpha$ (dashed lines in \Cref{fig:archnum}C), but even when there is ambiguity, p-values and significance levels have explicit interpretations as they are commonly used by practitioners for hypothesis testing.

\section*{Discussion}
PCA is commonly applied for dimension reduction. However, estimating the number of significant principal components that reflect the true signal in the data rather than noise can be challenging. Here, we propose a general approach to estimating the number of significant principal components that uses the posterior distribution inferred by VBPCA, a Bayesian version of the classical PCA algorithm \citep{ilin2010practical}. 

\begin{figure}[p]
    \centering
    \includegraphics[width=\textwidth]{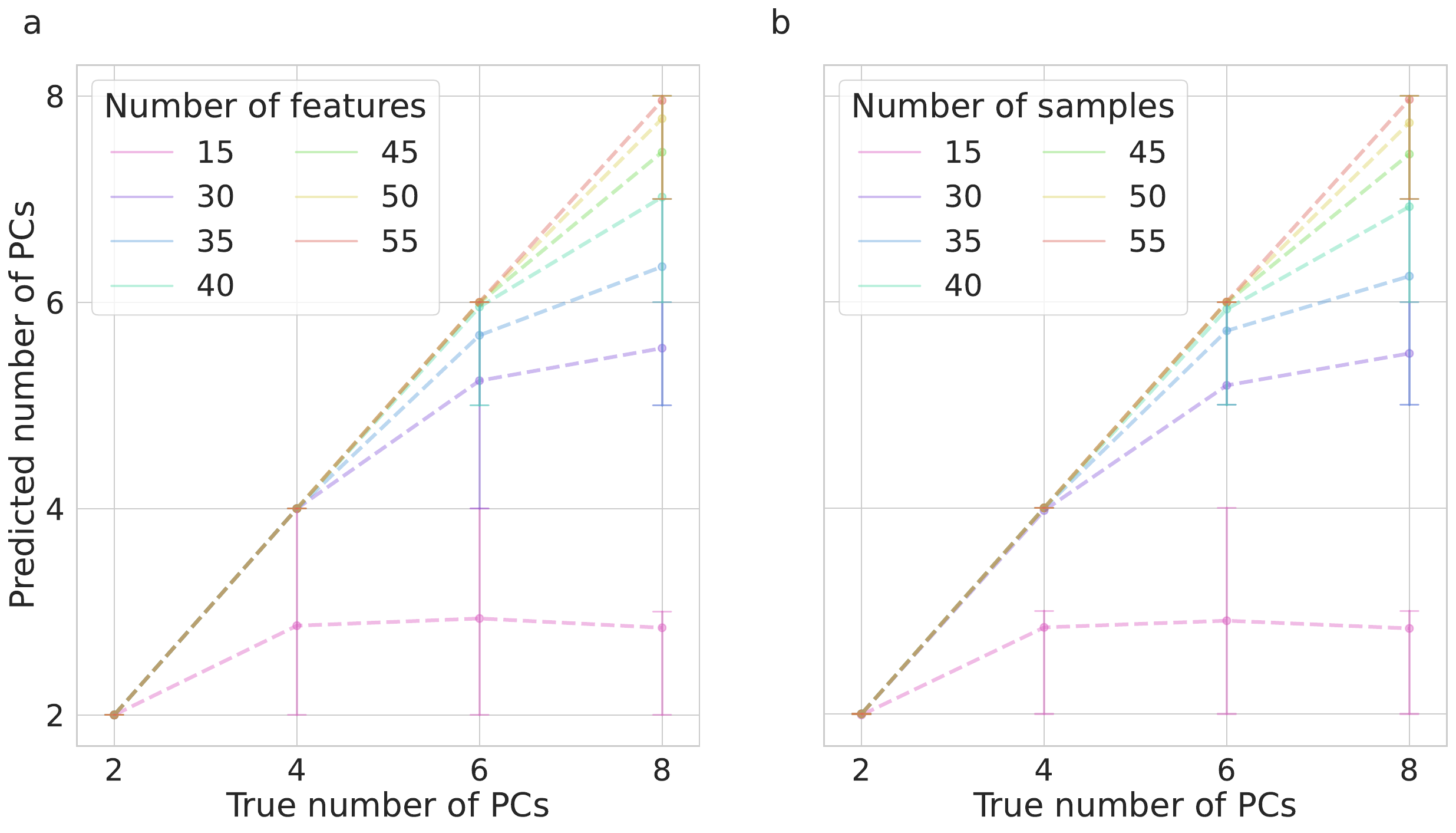}
    \caption{\textbf{Application to synthetic data.} 
    Our method sometimes under-estimated the number of significant PCs (y-axis), at least when the number of columns (panel a) or rows (panel b) is relatively small, but never over-estimated it.
    Here, we generated 200 synthetic datasets for scenario \textbf{(a)} with an increasing number of columns and 150 rows, and for scenario \textbf{(b)} with an increasing number of rows and 150 columns.
    Shown are the average estimated number of PCs (with 95\% confidence intervals) vs. the real number of PCs (which is known because we use synthetic data constructed to have a specific number of PCs).
    }
    \label{fig:testval}
\end{figure}

\begin{figure}[p]
\centering 
\includegraphics[width=0.8\textwidth]{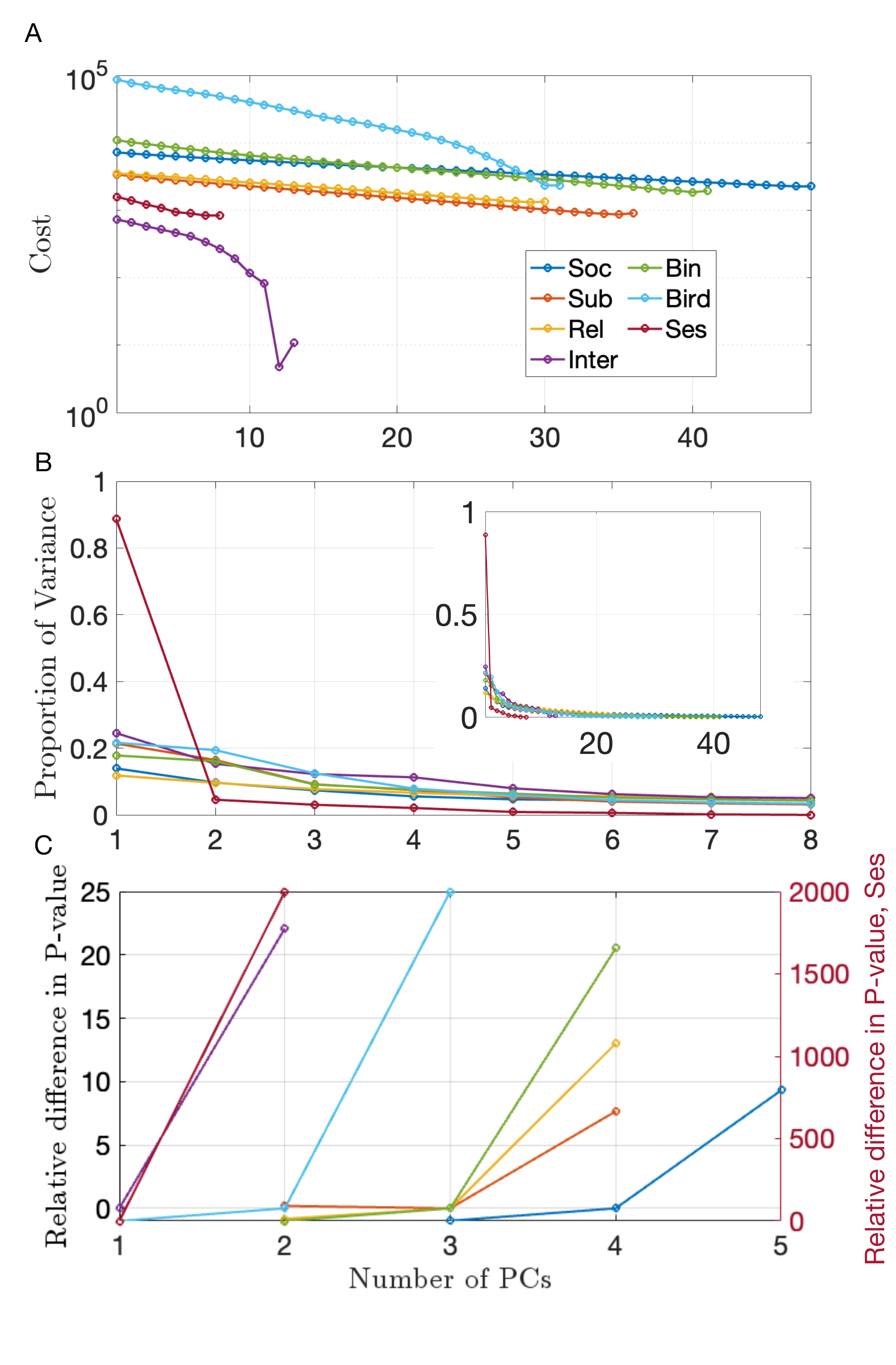}
\caption{ 
\textbf{Application to empirical data.}
\textbf{(A)} Number of principal components that optimally reconstruct each dataset (i.e., minimize the cost function, \cref{cost_func}). \textbf{(B)} Proportion of variance explained for each principal component. \textbf{(C)} Change in p-value around the estimated number of significant principal components, $w$. We show one less ($w-1$) and one more ($w+1$) than the estimated number ($w$) to show that the p-value "jump" from $w$ to $w+1$.
}
\label{fig:archnum}
\end{figure}

Our method is inspired by an approach suggested by \citet{patterson2006population} for population-genetic data, which requires that the number of columns is much larger than the number of rows.
We extend this approach to be applicable to datasets of any number of rows and columns.
Testing our new method on synthetic datasets, we find that although it sometimes under-estimates the number of significant principal components,  it does not appear to over-estimate this number. The method is quite accurate when the number of principal components is low, or the numbers of rows and columns are similar (\Cref{fig:testval}).
We also applied our method to empirical data: the cross-cultural datasets Pulotu, Ethnographic Atlas, Seshat, and Binford \citep{murdock1967ethnographic, pulotu,dplace,turchin2018quantitative}. 
We find evidence for a small number of significant principal components (between one and four), an order of magnitude less than the number of principal components required to optimally reconstruct the data (\Cref{fig:archnum}).
The estimated number of principal components can then be used for downstream analysis, such as clustering analysis \citep{Macdonald2023results}. 

Many previous studies have sought to estimate the number of significant principal components in various datasets \citep{kaiser1958varimax,cattell1966scree,patterson2006population,vasco2012permutation,courtney2013determining,karin2018tradeoffs,
turchin2018fitting,turchin2018quantitative,muthukrishna2020beyond,
beheim2021treatment,camargo2022pcatest}. 
For example, \cite{karin2018tradeoffs} previously analyzed datasets Rel and Inter (from Pulotu) in combination, ultimately performing Pareto task inference \citep{shoval2012evolutionary}. They used bootstrap to compare the variance explained by each principal component to the variance explained by a null model, which assumes that each trait is independent. They found two significant principal components by this approach. However, their null model is not based on distributions for each entry of the data matrix, as our test is. 
Another permutation approach is `exploratory factor analysis', where groups of columns (features) are correlated.
Both of the mentioned approaches require that missing data be handled via imputation or deletion prior to estimation of the number of principal components. As a result, depending on how the imputation is performed, the variance explained by a given number of principal components may be under- or over-estimated relative to VBPCA, which filters noise and imputes missing data \citep[Figure S20 in][]{Macdonald2023results}. 

`Cattell's scree test' \citep[][]{cattell1966scree} may be used when a few principal components describe an overwhelming proportion of variance. For example, \citet{turchin2018quantitative} analyzed a dataset where a single principal component explains the vast majority of the variance (Seshat in \Cref{fig:archnum} and \Cref{tab:results}).  
Our test agrees with the scree test in this case.
In the more common scenario, where the number of significant principal components cannot be easily determined by a scree test, our test uses the posterior distribution of the data to estimate the number of significant principal components. Thus, our test agrees with simple approaches in scenarios where tests for the number of significant components may not be strictly necessary. It provides a means to robustly estimate the number of significant components in cases where there is no clear estimate of the number of significant principal components.


{\small
\section*{\small Acknowledgements}
We thank Laura Fortunato, Sven Kasser, Daniel Weissman, Uri Obolski, Alex Ioannidis, and Tal Simon for their discussions and comments.
This research was supported in part by the John Templeton Foundation (MWF, YR), the Morrison Institute for Population and Resources Studies at Stanford University (MWF), the Minerva Stiftung Center for Lab Evolution (YR), and the Zuckerman STEM Leadership Program (JCM).
}

\clearpage



\clearpage

\section*{References}

\bibliographystyle{agsm}
\bibliography{MethodsPaper.bib}

@article{pulotu,
	author = {Watts, J and Sheehan, O and Greenhill, S J and Gomes-Ng, S and Atkinson, Q D and Bulbulia, J and others},
	journal = {PLoS ONE},
	number = {9},
	pages = {e0136783},
	title = {Pulotu: Database of {Austronesian} Supernatural Beliefs and Practices},
	volume = {10},
	year = {2015}}

@article{seshat,
	author = {Turchin, P. and others},
	journal = {Cliodynamics},
	number = {1},
	pages = {77-107},
	title = {Seshat: The Global History Databank},
	volume = {6},
	year = {2015}}

@article{dplace,
	author = {Kathryn R. Kirby and others},
	journal = {PLoS ONE},
	number = {7},
	pages = {e0158391},
	title = {D-PLACE: A Global Database of Cultural, Linguistic and Environmental Diversity},
	volume = {11},
	year = {2016}}

@article{tipping1999probabilistic,
	author = {Tipping, Michael E and Bishop, Christopher M},
	journal = {J Roy Stat Soc B Met},
	number = {3},
	pages = {611--622},
	publisher = {Wiley Online Library},
	title = {Probabilistic principal component analysis},
	volume = {61},
	year = {1999}}

@article{ilin2010practical,
	author = {Ilin, Alexander and Raiko, Tapani},
	journal = {The Journal of Machine Learning Research},
	pages = {1957--2000},
	publisher = {JMLR. org},
	title = {Practical approaches to principal component analysis in the presence of missing values},
	volume = {11},
	year = {2010}}

@misc{PPCAcode,
	author = {Alan Tran},
	title = {{PCA}-magic},
	url = {https://github.com/allentran/pca-magic},
	year = {2013}}

@misc{VBPCAcode,
	author = {Ilin, Alexander and Raiko, Tapani},
	title = {Matlab package for {PCA} for datasets with missing values},
	url = {https://users.ics.aalto.fi/alexilin/software/},
	year = {2010}}

@article{abdi2010holm,
	author = {Abdi, Herv{\'e}},
	journal = {Encyclopedia of Research Design},
	number = {8},
	pages = {1--8},
	publisher = {Thousand Oaks, California},
	title = {Holm's sequential {B}onferroni procedure},
	volume = {1},
	year = {2010}}

@article{patterson2006population,
	author = {Patterson, Nick and Price, Alkes L and Reich, David},
	journal = {PLoS genetics},
	number = {12},
	pages = {e190},
	publisher = {Public Library of Science San Francisco, USA},
	title = {Population structure and eigenanalysis},
	volume = {2},
	year = {2006}}

@article{karin2018tradeoffs,
	author = {Karin, Omer and Alon, Uri},
	journal = {bioRxiv},
	pages = {263905},
	publisher = {Cold Spring Harbor Laboratory},
	title = {Tradeoffs and cultural diversity},
	year = {2018}}

@article{shoval2012evolutionary,
	author = {Shoval, Oren and Sheftel, Hila and Shinar, Guy and Hart, Yuval and Ramote, Omer and Mayo, Avi and Dekel, Erez and Kavanagh, Kathryn and Alon, Uri},
	journal = {Science},
	number = {6085},
	pages = {1157--1160},
	publisher = {American Association for the Advancement of Science},
	title = {Evolutionary trade-offs, {P}areto optimality, and the geometry of phenotype space},
	volume = {336},
	year = {2012}}

@article{turchin2018quantitative,
	author = {Turchin, Peter and Currie, Thomas E and Whitehouse, Harvey and Fran{\c{c}}ois, Pieter and Feeney, Kevin and Mullins, Daniel and Hoyer, Daniel and Collins, Christina and Grohmann, Stephanie and Savage, Patrick and others},
	journal = {Proc Natl Acad Sci USA},
	number = {2},
	pages = {E144--E151},
	publisher = {National Acad Sciences},
	title = {Quantitative historical analysis uncovers a single dimension of complexity that structures global variation in human social organization},
	volume = {115},
	year = {2018}}

@article{murdock1967ethnographic,
	author = {Murdock, George Peter},
	journal = {Ethnology},
	number = {2},
	pages = {109--236},
	publisher = {JSTOR},
	title = {Ethnographic atlas: a summary},
	volume = {6},
	year = {1967}}

@article{turchin2018fitting,
  title={Fitting dynamic regression models to Seshat data},
  author={Turchin, Peter},
  journal={Cliodynamics},
  volume={9},
  number={1},
  year={2018}
}

@article{beheim2021treatment,
  title={Treatment of missing data determined conclusions regarding moralizing gods},
  author={Beheim, Bret and Atkinson, Quentin D and Bulbulia, Joseph and Gervais, Will and Gray, Russell D and Henrich, Joseph and Lang, Martin and Monroe, M Willis and Muthukrishna, Michael and Norenzayan, Ara and others},
  journal={Nature},
  volume={595},
  number={7866},
  pages={E29--E34},
  year={2021},
  publisher={Nature Publishing Group UK London}
}

@article{muthukrishna2020beyond,
  title={Beyond Western, Educated, Industrial, Rich, and Democratic (WEIRD) psychology: Measuring and mapping scales of cultural and psychological distance},
  author={Muthukrishna, Michael and Bell, Adrian V and Henrich, Joseph and Curtin, Cameron M and Gedranovich, Alexander and McInerney, Jason and Thue, Braden},
  journal={Psychological science},
  volume={31},
  number={6},
  pages={678--701},
  year={2020},
  publisher={Sage Publications Sage CA: Los Angeles, CA}
}

@article{cattell1966scree,
  title={The scree test for the number of factors},
  author={Cattell, Raymond B},
  journal={Multivariate behavioral research},
  volume={1},
  number={2},
  pages={245--276},
  year={1966},
  publisher={Taylor \& Francis}
}

@article{courtney2013determining,
  title={Determining the number of factors to retain in EFA: using the SPSS R-menu v2 0 to make more judicious estimations},
  author={Courtney, Matthew GR},
  journal={Practical Assessment, Research, and Evaluation},
  volume={18},
  number={1},
  pages={8},
  year={2013}
}

@article{camargo2022pcatest,
  title={PCAtest: testing the statistical significance of principal component analysis in R},
  author={Camargo, Arley},
  journal={PeerJ},
  volume={10},
  pages={e12967},
  year={2022},
  publisher={PeerJ Inc.}
}

@article{vasco2012permutation,
  title={Permutation tests to estimate significances on Principal Components Analysis},
  author={Vasco, MNCS},
  journal={Computational Ecology and Software},
  volume={2},
  number={2},
  pages={103},
  year={2012},
  publisher={International Academy of Ecology and Environmental Sciences (IAEES)}
}

@article{kaiser1958varimax,
  title={The varimax criterion for analytic rotation in factor analysis},
  author={Kaiser, Henry F},
  journal={Psychometrika},
  volume={23},
  number={3},
  pages={187--200},
  year={1958},
  publisher={Springer}
}

@article{Macdonald2023results,
  title={Cultural transmission, networks, and clusters among Austronesian-speaking peoples},
  author={Macdonald, Joshua C and Blanco-Portillo, Javier and Feldman, Marcus W and Ram, Yoav},
  journal={Evolutionary Human Sciences},
  volume={},
  number={},
  pages={},
  year={In Press},
  publisher={}
}

@book{binford2019constructing,
  title={Constructing frames of reference: an analytical method for archaeological theory building using ethnographic and environmental data sets},
  author={Binford, Lewis R},
  year={2019},
  publisher={University of California Press}
}

@article{tobias2022avonet,
  title={AVONET: morphological, ecological and geographical data for all birds},
  author={Tobias, Joseph A and Sheard, Catherine and Pigot, Alex L and Devenish, Adam JM and Yang, Jingyi and Sayol, Ferran and Neate-Clegg, Montague HC and Alioravainen, Nico and Weeks, Thomas L and Barber, Robert A and others},
  journal={Ecology Letters},
  volume={25},
  number={3},
  pages={581--597},
  year={2022},
  publisher={Wiley Online Library}
}

\end{document}